\documentclass[12pt]{article}
\usepackage{epsf}
\begin{document}
\begin{center}
{\large\bf Implementation of the Hough Transform\\ 
           for 3D Track Reconstruction in Drift Chambers\\
}
\vspace*{3mm}
{\bf Ar.Belkov}\\[6pt]
Laboratory of Particle Physics, JINR\\
\vspace*{3mm}
{\it Talk at the VIth International School-Seminar\\ 
      ``Actual Problems of High Energy Physics''\\
     August 7-16, 2001, Gomel, Belarus}
\end{center}
\vspace*{2mm}

   This paper is devoted to the method developed in Ref.~\cite{belkov} for 
3D reconstruction of the straight tracks in the tracking system consisting 
of the drift-chamber stereo layers.
   The method is based on the Hough-transform approach \cite{hough} -- 
the discrete case of more general Radon transform \cite{radon} -- and takes 
into account both coordinates of the hit wires and drift distances not only 
for the measurements in one projection, but also in the rotated stereo layers.
   The proposed method allows one to resolve the right-left ambiguity and 
provides the accordance between vertical and horizontal projections of the 
track.

   Let the straight track of charged particle be detected by the system of 
{\it drift chambers} consisting of the cylindrical tubes placed in such a 
way that their anode wires stretched along the tube axes are parallel to the 
vertical coordinate axis $Y$.
   In this case the signals from the drift chambers give the information about
the track projection onto the horizontal plane $XZ$.
   The track projection is described by $x(z)= x_0+t_x(z-z_0)$, where $x_0$ 
and $t_x=\mbox{tan}\,\theta_x$ are the projection offset in $z$ and slope, 
respectively.

\begin{figure}[hbt]
\epsfxsize=135mm\epsfbox{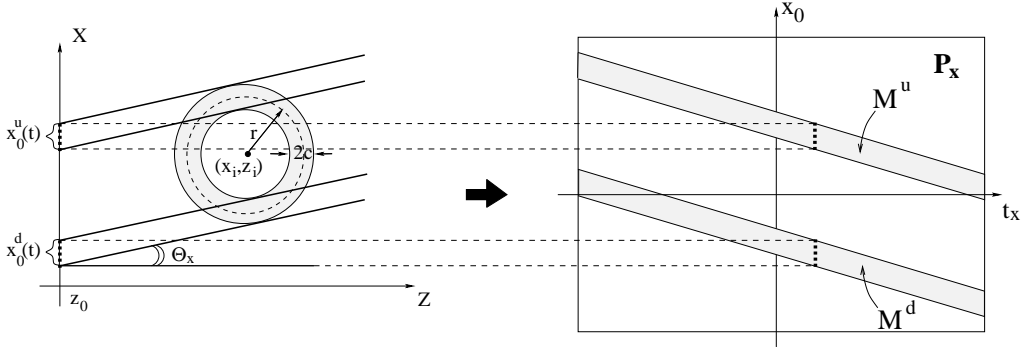}
\caption{\small The Hough image of the single measurement $(x_i, z_i, r_i)$ 
                in the space $P_x$ of projection parameters $(x_0, t_x)$}
\label{fig1}
\end{figure}

   {\it The single measurement} from the hit wire includes its coordinates 
$x_i$, $z_i$ and measured distance $r_i$ from the wire to the track.
   Assume that error of $r_i$ measurement is uniformly distributed within the 
range $[-c, c]$, where $c \approx (2 \div 3) \sigma$ is a tuning parameter of 
the algorithm while $\sigma$ is a space resolution of the drift chamber.
   At each value of $t_x$, {\it the Hough image of a single measurement}
$(x_i, z_i, r_i)$ in the space $P_x$ of projection parameters $(x_0, t_x)$
is given by two ranges of possible values of $x_0$ 
(see Fig.~\ref{fig1}):
\vspace*{-1mm}
\begin{equation}
x^u_0 \in \Big[x_i-(z_i-z_0)t_x + (r_i-c) \sqrt{1+t^2_x}\,;\,
               x_i-(z_i-z_0)t_x + (r_i+c) \sqrt{1+t^2_x}\Big]
\label{ref1}
\end{equation}
is an upper branch $M^u$,
\vspace*{-2mm}
\begin{equation}
x^d_0 \in \Big[x_i-(z_i-z_0)t_x - (r_i+c) \sqrt{1+t^2_x}\,;\,
               x_i-(z_i-z_0)t_x - (r_i-c) \sqrt{1+t^2_x}\Big]
\label{ref2}
\end{equation}
is a lower branch $M^d$.

   For chamber operation in the {\it proportional regime} (no measurement of
$r_i$), the Hough image of a single hit $(x_i, z_i)$ is given in the 
parameter space $P_x$ by a range of possible values of $x_0$ at each value of 
$t_x$:
\vspace*{-1mm}
\begin{equation}
x_0 \in \Big[x_i-(z_i-z_0)t_x - R \sqrt{1+t^2_x}\,;\,
             x_i-(z_i-z_0)t_x + R \sqrt{1+t^2_x}\Big]\,,
\label{ref3}
\end{equation}
where $R$ is the tube radius.

   Assume $M_1, M_2, \ldots, M_n$ -- are Hough images of single measurements
from $n$ hits belonging to the same track.
   In this case {\it the Hough image of the track}, $M^{(n)}$, is defined in 
the parameter space $P_x$ as an intersection of set of Hough images ${M_k}$, 
$k=1,\ldots,n$: $M^{(n)}= \bigcap^n_{k=1} M_k$.
   To characterize the track reliability level, let us introduce the criterion 
$J(M)$ which can be determined as number of hits having produced the Hough 
image $M$ of the track.
   In particular, $J(M^{(n)})=n$.
   For the reconstruction we use only those $M$ for which $J(M) \geq J_{min}$,
where a threshold value of $J_{min}$ is a tuning parameter of the algorithm.
   If the intersection $M$ is found, then the track parameters $x_0$, $t_x$ 
are estimated as coordinates of center of gravity of $M$ in the parameter 
space $P_x$.

\begin{figure}[hbt]
\hspace*{25mm}
\epsfxsize=90mm\epsfbox{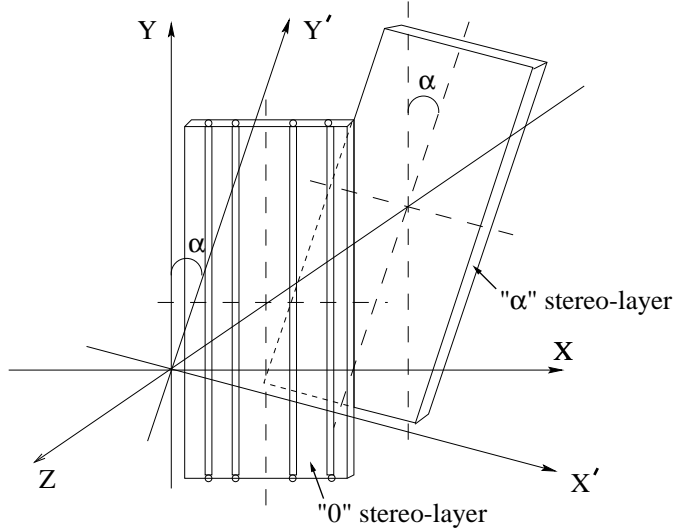}
\caption{\small Stereo-layers in the $XYZ$ and rotated  
                $X^{\prime}Y^{\prime}Z$ coordinate systems}
\label{fig2}
\end{figure}

   For 3D reconstruction, it is also necessary to determine the track 
projection onto the vertical plane $YZ$.
   For this aim, the set of vertical tubes ({\it ``0'' stereo-layers}) with 
set of tubes rotated by the angle $\alpha$ around the axis $Z$ 
({\it ``$\alpha$'' stereo-layers}) can be used as it is shown in 
Fig.~\ref{fig2}.
   The $YZ$ track projection is described by $y(z)=y_0+t_y(z-z_0)$, 
where $y_0$ and $t_y=\mbox{tan}\theta_y$ are the projection offset in $z$ and 
slope, respectively.

   Assume parameters $x_0$, $t_x$ for the $XZ$ track projection have already 
been determined using measurements in the ``0'' stereo-layers and 
Hough-transform approach described above.
   Then,~ the single measurements $(x^{\prime}_i, z_i, r_i)$ in ``$\alpha$'' 
stereo-layers with hit-wire coordinates determined in the rotated system 
$X^{\prime}Y^{\prime}Z$ (see Fig.~\ref{fig2}) can be used for reconstruction 
of track projection onto the plane $YZ$. 
   The corresponding Hough images of single measurements are defined
in the space $P_y$ of the parameters $y_0$, $t_y$ by the following boundaries
of possible values of $y_0$ at fixed $t_y$ (see Ref.~\cite{belkov} for more 
detail):
\vspace*{-5mm}
\begin{eqnarray}
y_0(t_y; x_0, t_x)&=& \frac{x_0+t_x(z_i-z_0)}{\mbox{tan}\,\alpha}
                     -\frac{x^{\prime}_i}{\mbox{sin}\,\alpha}-t_y(z_i-z_0)
\nonumber\\
		&\pm& \frac{r_i\pm c}{\mbox{sin}\,\alpha}
                      \sqrt{1+(t_x\mbox{cos}\,\alpha-t_y\mbox{sin}\,\alpha)^2}
\label{ref4}
\end{eqnarray}
-- for the drift chambers, or
\vspace*{-4mm}
\begin{eqnarray}
y_0(t_y; x_0, t_x)&=& \frac{x_0+t_x(z_i-z_0)}{\mbox{tan}\,\alpha}
                     -\frac{x^{\prime}_i}{\mbox{sin}\,\alpha}-t_y(z_i-z_0)
\nonumber\\
		&\pm& \frac{R}{\mbox{sin}\,\alpha}
                      \sqrt{1+(t_x\mbox{cos}\,\alpha-t_y\mbox{sin}\,\alpha)^2}
\label{ref5}
\end{eqnarray}
- for chamber operation in proportional regime.

   For the program realization of the discussed algorithm, for example, for
$XZ$-projection finding, the space $P_x$ of the track parameters $x_0$, $t_x$ 
can be treated as a {\it discrete two-dimensional raster} $n\times m$, which 
is described by array of its cells $\rho(l,k)$ with indices $l=1,\ldots,n$ and
$k=1,\ldots,m$.
   The Hough image of each hit can be consequently constructed as a strip on 
the raster: the current value of the raster cell is increased by a unit if the 
cell is located within the limits given by Eqs.~(\ref{ref1}), (\ref{ref2}) or 
(\ref{ref3}).
   After completing this procedure the value of each cell of the raster
becomes equal to the number of the Hough stripes having passed the cell.
   Each local maximum on the raster exceeding some threshold $J_{min}$, can be
identified with the $XZ$ projection having the corresponding values of 
parameters $x_0$, $t_x$.

   The selection algorithm of reliable tracks includes ordering the raster 
cells according to the criterion $J(M)$ by using a special structure -- the
{\it array of bidirectional lists} determining the {\it hierarchy of the 
raster cells} filled in.
   The given array of the lists is a vector-column where each element is
a pointer to the list and corresponds to a certain meaning of the criterion 
$J(M) > J_{min}$.
   After filling the raster, the hierarchy of its cells is built: if 
$\rho(l,k)\geq J_{min}$, then the point $(l, k)$ is added into the list for 
$J=\rho(l, k)$.

   At the first step of 3D track reconstruction, the global $XZ$ raster 
and its hierarchy are built by using the hits only in the ``0'' stereo-layers. 
   The further steps of the algorithm include:
\begin{itemize}
\vspace*{-2mm}
\item estimating of the initial values of $x_0$, $t_x$ corresponding 
      to the first maximum in the global $XZ$-raster hierarchy;
\vspace*{-2mm}
\item building the local $XZ$ raster and its hierarchy by using the hits 
      in the ``0'' stereo-layers within a corridor around the track projection 
      defined by the initial values of $x_0$, $t_x$;
\vspace*{-2mm}
\item improvement of the estimates of $x_0$, $t_x$ as coordinates of center of 
      gravity of the local-raster maximum;
\vspace*{-2mm}
\item building the $YZ$ raster and its hierarchy assigned to the found $XZ$ 
      projection $(x_0, t_x)$, by using hits only in the ``$\alpha$'' 
      stereo-layers;
\vspace*{-2mm}
\item finding of $YZ$ projection $(y_0, t_y)$ assigned to the found $XZ$
      projection $(x_0, t_x)$, by using the $YZ$-raster hierarchy;
\vspace*{-2mm}
\item subtraction of the hits in ``0'' and ``$\alpha$'' stereo-layers, which
      belong to the 3D-track $(x_0, t_x; y_0, t_y)$, and consecutive
      subtraction of the corresponding Hough strips from the global raster
      and its hierarchy.
\end{itemize}
\vspace*{-2mm}
   The procedure described above should be repeated iteratively until there 
remain the cells of the global $XZ$ raster exceeding some threshold $J_{min}$.

\begin{figure}[hbt]
\begin{center}
\begin{tabular}{cc}
\hspace*{-6mm}\epsfxsize = 0.50\textwidth\epsfbox{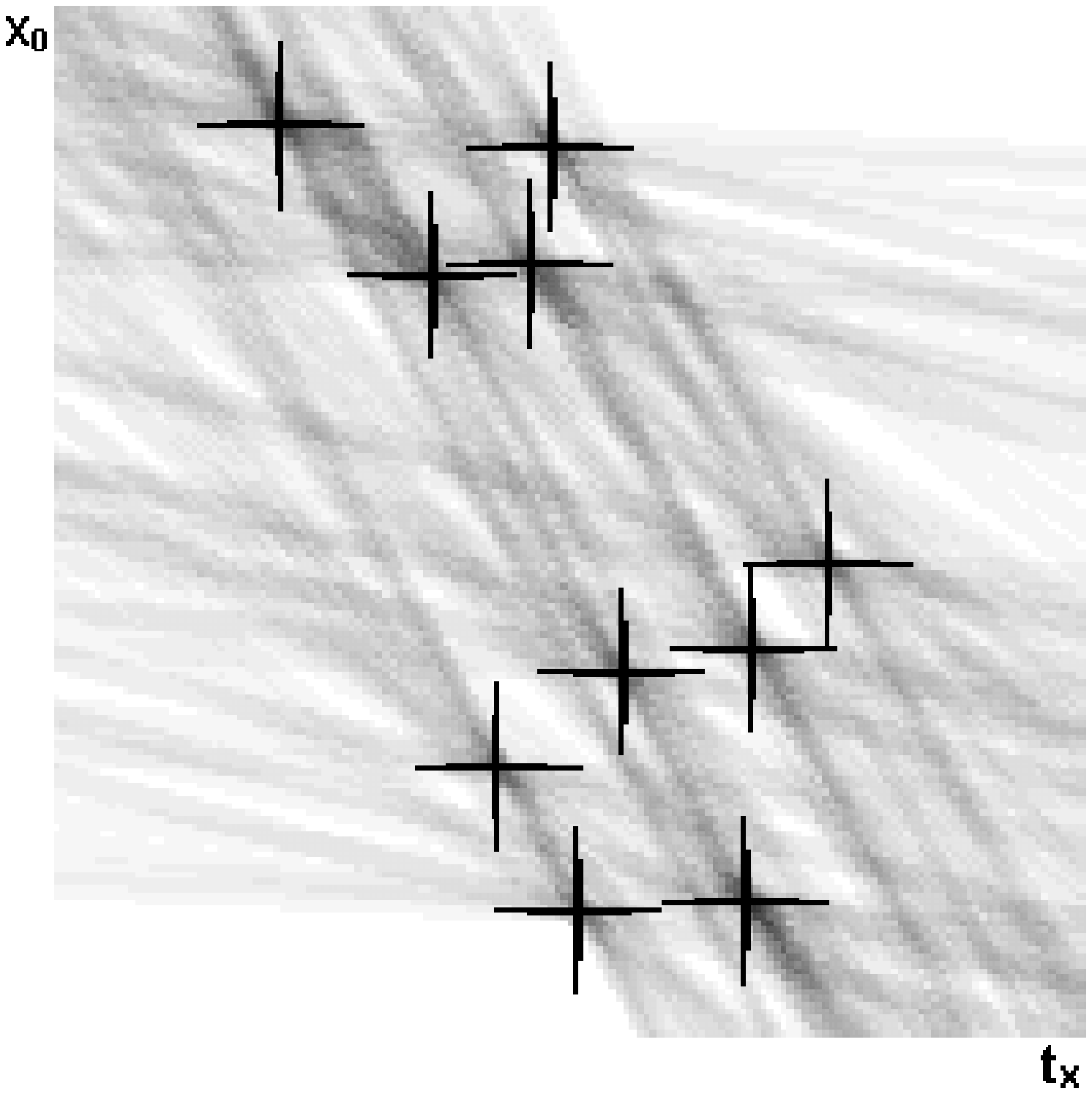} &
              \epsfxsize = 0.50\textwidth\epsfbox{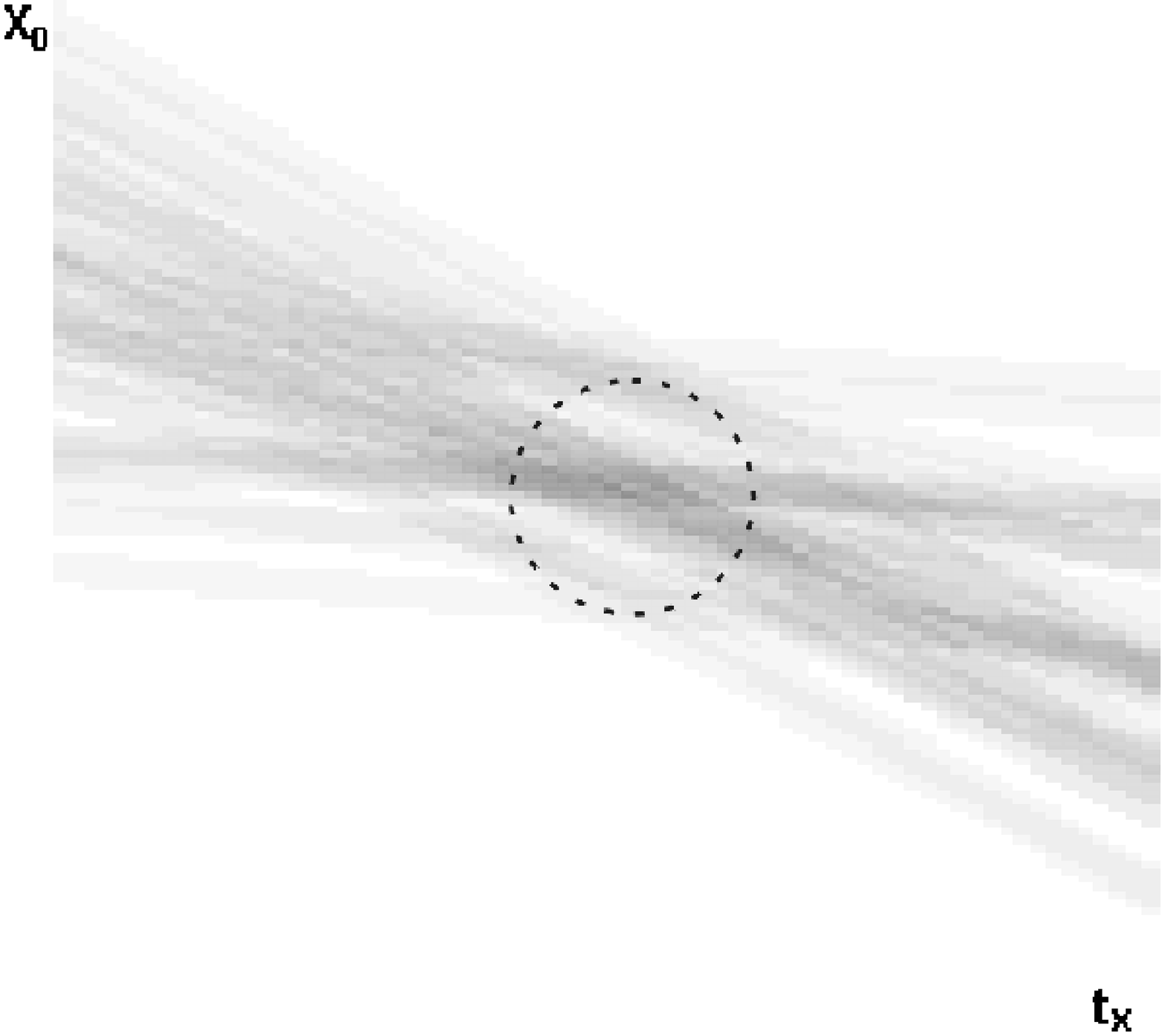}\\
  a) & b)

\end{tabular}
\caption{\small {Examples of global (a) and local (b) $XZ$ rasters}}
\vspace*{-5mm}
\end{center}
\label{fig3}
\end{figure}

   The examples of global and local $XZ$ rasters for Monte-Carlo 
tracks are shown in Fig.~3.
   The small crosses on the global raster mark Monte-Carlo tracks
on the ($x_0, t_x$)-plane of the track projection parameters while the large
crosses correspond to the values $x_0$, $t_x$ of reconstructed projections.
   Fig.~3a shows a good precision of track reconstruction by the 
Hough-transform algorithm described above.
 
   This algorithm has been realized in the program {\it Htr} developed for
track finding in the PC chambers (Pattern Tracker) of the HERA-B Outer Tracker 
\cite{HERA-B}.
   The program {\it Htr} was integrated into the program environment of 
{\it ARTE} -- the general software for event processing at HERA-B -- and 
tested both with Monte-Carlo and real data.
   The tests showed the stable {\it Htr} performance with average track
finding efficiency of about 90\% and rate of ghosts at the level 23\%
under real conditions of PC-chamber operation.
   The proposed program realization of the described algorithm provides 
the time-consuming optimization of event processing and high efficiency of 
the track finding under large track-occupancy of the detector as well 
as under high level of noisy and dead channels.

   This work was done in the HERA-B software group at DESY, Hamburg.

\vspace*{-5mm}

\end{document}